\documentclass[conference]{IEEEtran}
\IEEEoverridecommandlockouts    

\usepackage{multirow}
\usepackage{lipsum}
\usepackage{amsmath}
\usepackage{amssymb}
\usepackage[ruled,vlined]{algorithm2e}
\usepackage{graphicx}
\usepackage{booktabs}
\usepackage{siunitx} 
\usepackage{makecell}
\usepackage{orcidlink}
\usepackage{tabularx}
\usepackage{array}
\usepackage{xcolor}
\usepackage{comment}
\usepackage{todonotes}

\usepackage[T1]{fontenc}

\usepackage{subcaption}
\usepackage{orcidlink}
\usepackage{hyperref}
\hypersetup{
    colorlinks=true,
    linkcolor=blue,
    citecolor=blue,
    filecolor=magenta,      
    urlcolor=blue
}

\graphicspath{{./Figures/}}

\begin{document}

\title{Evaluating the Health of\\Open-Source Smart City Platforms

\thanks{This research was partially supported by national funds through 
Fundação para a Ciência e a Tecnologia, I.P. (FCT), under Pluriannual Projects \href{https://doi.org/10.54499/UID/04466/2025}{UID/04466/2025} (ISTAR-Iscte) and \href{https://doi.org/10.54499/UID/04516/2025}{UID/04516/2025} (NOVA LINCS).}
}


\author{
    \IEEEauthorblockN{%
        Rodrigo Bravo Simões\IEEEauthorrefmark{1},
        Fernando Brito e Abreu\IEEEauthorrefmark{1},
        Vasco Amaral\IEEEauthorrefmark{2}
    }
    \IEEEauthorblockA{%
        \IEEEauthorrefmark{1}Instituto Universitário de Lisboa (ISCTE-IUL), ISTAR, Lisboa, Portugal\\
        \{rjbss,fba\}@iscte-iul.pt
    }
    \IEEEauthorblockA{%
        \IEEEauthorrefmark{2}NOVA LINCS, DI, FCT/UNL, Lisboa, Portugal\\
        vma@fct.unl.pt
    }
}

\maketitle

\begin{abstract}
    To manage the complexity of smart cities, a variety of smart city platforms (SCPs), both proprietary and open-source, have been proposed. Typically acting as middleware between IoT devices at a lower layer and smart services at a higher layer, these platforms simplify the management of smart cities by reifying a variety of requirements common to different municipalities. Open-source software (OSS) is typically free of charge, making open-source SCPs an attractive option for municipalities with smaller budgets that still wish to improve efficiency and quality of life for residents and visitors. Furthermore, using an OSS SCP promotes digital sovereignty, since municipalities can control where the data will reside and will be less prone to vendor lock-in.    
    
    A lack of support for OSS hinders its adoption by municipalities' IT teams, often with scarce technical resources. An active (aka ``healthy'') ecosystem, including well-orchestrated developers and users, the availability of installation and user manuals, non-vulnerable and documented code, test batteries, and other artifacts, is an important decision factor before adopting an OSS SCP. In this paper, we evaluate different open-source SCPs using the concept of ``OSS health'', covering several of the aforementioned characteristics.
\end{abstract}


\section{Introduction}
\label{sec:introduction}

While the transformation of a city into a smart city has different specificities depending on the concrete reality of each region, many requirements are common between smart cities, such as the usage of information and communication technology (ICT) and, more specifically, IoT (Internet of Things) devices, to manage resources more efficiently and improve the quality of urban life~\cite{Silva2018}. One popular strategy to cope with the complexity of the software infrastructure that is required is to deploy a smart city platform (SCP), which typically acts as a middleware between IoT devices at a lower layer and smart services at a higher layer.  

While each municipality can create its own SCP to deal with its specific use cases, the commonalities between smart cities can be exploited to implement generic SCPs that are reusable across different cities. Releasing an SCP as open-source software (OSS) arises as an opportunity to collectively improve the software infrastructure of our municipalities. An example of this is \textit{FIWARE}, which claims to be running in more than 400 cities~\cite{FIWARE_book}. OSS is typically free of charge, making open-source SCPs an attractive option for municipalities with smaller budgets. Furthermore, using an OSS SCP promotes digital sovereignty, since municipalities can control where the data will reside and will be less prone to vendor lock-in.  

Lack of support for OSS is one critical factor that hinders adoption in public organizations, including local governments~\cite{Yaseen2024}. Consequently, assessing OSS health~\cite{Linaaker2022} is a crucial step when comparing open-source SCPs, as an active community often serves as the primary support mechanism. A related challenge is the typical shortage of specialized technical staff in smaller municipalities. As SCPs usually require significant technical knowledge for deployment and maintenance, high-quality documentation arises as a particularly important characteristic of these projects' health, as it helps smaller municipal IT teams to successfully deploy and operate the platform.

Thus, in this paper we analyze the health of six open-source SCPs, mostly based on metrics collected from their source code repositories. Special emphasis is given to the quality of the documentation, the analysis of which is greatly aided by large language models (LLMs).

The remainder of this paper is as follows: in Section~\ref{sec:related_work} we look at prior work in SCP comparison and OSS health evaluation; in Section~\ref{sec:analysis} we present the methodological choices, intertwined with their corresponding results; then, in Section~\ref{sec:discussion} we consolidate and discuss the results presented in the previous section; finally, in Section~\ref{sec:conclusion_future_work} we conclude the paper and outline future work.

\section{Related Work}
\label{sec:related_work}

Multiple secondary studies have been written about SCPs. One of these has been presented in an earlier iteration of this conference~\cite{Achilleos2019}. It compares open-source SCPs, based on a list of requirements that was extracted from earlier works, performing a comparative analysis based on those requirements. A more complete analysis of SCP requirements, along with descriptions of the technologies that enable them, has been done by Goumopoulos~\cite{Goumopoulos2024} through a synthesis of available information about various SCPs. There is an earlier survey paper with similar goals that additionally presents a reference architecture for SCPs~\cite{Santana2017}.

With solid prior work on requirements for SCPs, there is another dimension that, to the best of our knowledge, has not been explored in this domain: OSS health. A literature review of OSS health~\cite{Linaaker2022} has extracted 107 health characteristics divided among 15 themes. It does not prescribe a methodology for selecting which characteristics to evaluate, unlike many of its primary sources. Similarly, the \href{https://www.chaoss.community/}{CHAOSS project}, which focuses on metrics for OSS community health, does not prescribe such a framework; instead, the authors recommend devising user stories to express the goals of the health analysis, from which the relevant metrics can be derived~\cite{Goggins2021}. Our work consists of obtaining metrics from GitHub and GitLab repositories of active SCPs, using \href{https://github.com/aveloxis/aveloxis}{Aveloxis}, a tool that supersedes \href{https://github.com/augurlabs/augur}{Augur}; the latter was presented in the CHAOSS paper~\cite{Goggins2021}.

Another related work evaluated open-source IoT platforms hosted on GitHub~\cite{Turki2024}. The authors compared the repositories based on activity metrics and popularity indicators. Since we were empowered by Aveloxis, we could gather a wider variety of metrics that go beyond what is available in GitHub's API. On the other hand, we do not go into as much detail about the features of the various platforms. We note that the concept of an IoT platform is close to that of an SCP, as a smart city can be seen as an application of the IoT concept~\cite{Silva2018}. Indeed, the previously cited study on IoT platforms~\cite{Turki2024} includes one platform that is also assessed in our paper (\textit{OpenRemote}). The boundaries of what constitutes an SCP can vary by author; readers interested in this issue can refer to a work in which a taxonomy of SCPs is presented~\cite{Richter2023}.

\section{Analysis}
\label{sec:analysis}

\subsection{Gathering Repositories}

We gathered a list of open-source SCP repository sources from the survey papers on SCPs mentioned in Section~\ref{sec:related_work}~\cite{Santana2017,Achilleos2019,Goumopoulos2024} and a technical report~\cite{Weber2025}. Firstly, we compiled the names of the SCPs listed in these sources, excluding those without a name (typical of academic prototypes not yet mature enough for real-world adoption). For each named platform, we conducted a Google search. For an SCP named X, the search string was \texttt{"X" AND (source OR repository)}. If the name had other uses and the first page of search results contained only irrelevant entries, the search string was augmented with \texttt{AND ("smart city" OR IoT)}.

This approach yielded an initial set of fifteen GitHub and GitLab entries, comprising individual repositories and organizational groups. For the groups, we selected the single-core repository representing the project. We then filtered the resulting 15 repositories, excluding those with no commits or only a nominal number of trivial commits over the past two years. This process resulted in a final selection of six repositories.

\textit{FIWARE}, one of the organizational groups on GitHub, does not have an obvious single core repository that represents the project. Instead, it provides an extensive catalog of components called \href{https://www.fiware.org/catalogue/}{\textit{FIWARE Generic Enablers}}. For our analysis, we chose one of the four available context brokers because a context broker component is the core and mandatory component of any ``Powered by FIWARE'' platform or solution. These brokers can mediate communication between devices spread around the city and the smart services that use these devices. In short, they can serve as the core of an SCP. We selected the \textit{Orion-LD} context broker to represent \textit{FIWARE}, since it is the only context broker that is developed by the FIWARE Foundation itself.



\subsection{Overview}
\label{subsec:overview}

\begin{table*}[t]
    \centering
    \small
    \caption{Descriptive characteristics of the smart city platforms included in this study (alphabetic order)}
    \label{tab:descriptive}
    
    \begin{tabular}{
        >{\raggedright\arraybackslash}p{5cm}
        >{\raggedright\arraybackslash}p{2cm}
        >{\raggedright\arraybackslash}p{4cm}
        >{\raggedright\arraybackslash}p{1.5cm}
        p{1.5cm}
        l
    }
        \toprule
        
        \textbf{Platform codebase}      &
        \textbf{Platform site}      &
        \textbf{Producer / Maintainer}        &
        \textbf{Origin}        &
        \textbf{Language}        &
        \textbf{License}        \\
        
        \midrule
        
        \href{https://gitlab.com/civitas-connect/civitas-core/civitas-core}{CIVITAS/CORE}   &
        \href{https://www.civitasconnect.digital/civitas-core/}{link} &
        \href{https://civitasconnect.digital/}{Civitas Connect e.V.}        &
        Germany       &
        Python        &
        EUPL v1.2     \\
        
        \href{https://github.com/smartfog/fogflow}{FogFlow}     &
        \href{https://fogflow.readthedocs.io/en/latest/#}{link} &
        \href{https://neclab.eu/}{NEC Laboratories Europe}      &
        EU        &
        Go        &
        BSD 3-Clause     \\
        
        \href{https://github.com/openremote/openremote}{OpenRemote}  &
        \href{https://openremote.io/}{link} &
        \href{https://openremote.io/}{OpenRemote B.V.}      &
        Netherlands      &
        Java             &
        GNU AGPL         \\
        
        \href{https://github.com/FIWARE/context.Orion-LD}{Orion-LD}      &
        \href{https://fiware-orion.readthedocs.io/en/master/index.html}{link}  &
        \href{https://fiware.org/}{FIWARE Foundation e.V.}  &
        Germany       &
        C++           &
        GNU AGPL      \\
        
        \href{https://github.com/disit/snap4city}{Snap4City}         &
        \href{https://www.snap4city.org/}{link}                      &
        \href{https://www.disit.org/}{DISIT Lab, Univ. of Florence}  &
        Italy        &
        JavaScript   &
        GNU AGPL     \\
        
        \href{https://gitlab.com/urban-dataspace-platform/core-platform}{Urban Dataspace Platform (UDP)}   &
        \href{https://www.hypertegrity.de/urban-data-space-platform/}{link}   &
        \href{https://www.hypertegrity.de/}{Hypertegrity AG}   &
        Germany      &
        Jinja        &
        EUPL v1.2    \\
        
        \bottomrule
    \end{tabular}
\end{table*}

Table~\ref{tab:descriptive} lists the repositories whose analysis we will present in the following subsections. The table includes the SCP name with a link to the repository, its producer organization and the respective country/region of origin, the number of stars at the time of writing (in GitHub or GitLab), the number of commits on the main branch in the two years between 2024-06-01 and 2026-06-01, the total number of lines of code in the latest commit of the main branch, and the main language present in the repository. The following paragraphs provide additional information to give the reader a clearer picture of the SCPs involved. This information was manually compiled based on the official documentation for each platform.

\textit{CIVITAS/CORE} is a fork of the \textit{Urban Dataspace Platform (UDP)}, which is also covered by our analysis. It has been developed by a consortium comprising nine cities and regions, as well as eight municipal companies. The platform reuses many components, including \textit{FIWARE Generic Enablers}, to provide a pre-packaged solution which is easier to install and configure than assembling the same components manually. Although Python is the primary programming language in the repository, most Python files are for testing purposes only, while declarative YAML files do the heavy lifting.

\textit{FogFlow} is a \textit{FIWARE Generic Enabler} that serves as a cloud-edge orchestration framework. While it is not a complete SCP on its own, it is ready to be integrated with other components to create one. These components include an NGSI-LD context broker, such as \textit{Orion-LD}, which is also covered by our analysis. \textit{FogFlow} provides a programming model and development tools that enable the rapid implementation of IoT services and smart city applications with optimized quality of service (QoS).

\textit{OpenRemote} is an IoT platform that can serve as a centralized system for smart cities. It provides device integration, rule-based automation, analytics, and natively supports several protocols via agents. The \textit{OpenRemote} website lists several smart city-specific use cases that the platform can facilitate, including data integration, traffic flow prediction, automated notifications, and dashboard apps. The platform includes front-end user interface web components to facilitate the development of custom applications for end users, as well as a dashboard builder to further accelerate development.

\textit{Orion-LD} is a fork of the Orion context broker, adding support for NGSI-LD. Like the previous version of NGSI (NGSI-v2), NGSI-LD enables the management and exchange of context information for smart technologies, but is augmented with linked data capabilities. \textit{Orion-LD} is thus a suitable candidate for the core component of an SCP. Other key features of NGSI-LD that are particularly helpful in the development of smart cities include the ability to create temporal representations of entities and support for geographical data. Furthermore, the adoption of the linked data paradigm has the potential to improve interoperability across different platforms.

\textit{Snap4City} is a platform based on the \textit{Km4City Ontology} \cite{Bellini2014} for its knowledge base and on the Smart City API. It provides a development environment based on Node-RED and JavaScript that supports many different IoT protocols and types of devices. Several smart city tools have been developed using it across various smart city vertical domains. For example, the platform's website offers an extensive set of dashboards that exploit public data. In addition to providing analytics and decision support, the platform aims to act as an AI-enabled digital twin. Its development platform contains a set of ready-to-use AI modules to facilitate the creation of smart services.

Similar to \textit{CIVITAS/CORE}, \textit{UDP} provides an integration of various components with the goal of realizing an SCP. Its current architecture contains tens of these components, which are organized into various stacks, including identity, context, API, and data management, among others. Similar to \textit{CIVITAS/CORE}, most of its codebase consists of YAML configuration files for the various components, and the templating language of the \href{https://jinja.palletsprojects.com/en/stable/}{Jinja templating engine} is frequently used.

\subsection{Open-source licenses}

Software licensing is an important consideration when selecting an open-source smart city platform, as it directly affects deployment, customization, and redistribution. As shown in Table~\ref{tab:descriptive}, \textit{FogFlow} is the only platform released under a permissive license (BSD 3-Clause), whereas the remaining platforms adopt copyleft licenses, namely \href{https://www.gnu.org/licenses/why-affero-gpl.html}{GNU AGPL} or \href{https://eupl.eu/1.2/en/}{EUPL v1.2}. Although both licenses require the publication of derivative works under certain conditions, the EUPL follows a variable copyleft model, providing greater compatibility with other licenses and potentially allowing integration with proprietary software under specific circumstances~\cite{Schmitz2013}. By contrast, the AGPL enforces strong copyleft, requiring source code modifications to be made available even when the software is offered exclusively as a network service. Given the diversity of licensing schemes, municipalities intending to customize and deploy open-source platforms should carefully assess their legal obligations and seek appropriate legal guidance before operational deployment.

\subsection{Documentation Quality}

To analyze the quality of the documentation, we devised a group of 18 binary Yes/No questions that were to be answered by reading the documentation. These questions target 21 of the various quality aspects enumerated by Tang in a study about the evaluation of documentation quality~\cite{Tang2023}. Note that some of these questions target more than one quality aspect. These questions were designed not only to target a diverse set of quality aspects, but also to target different practical issues that arise when configuring and using an SCP, such as prerequisites, deployment, and interoperability. The questions are available in the \href{https://zenodo.org/records/22646494}{supplementary material}.

We fed the platforms' documentation to three different Large Language Models (LLMs), via \href{https://openrouter.ai/}{OpenRouter}, a service that routes requests to models of different authors and providers. For each platform, we created a prompt consisting of instructions, questions, and the documentation. Since the documentation files were quite big (reaching more than 1MB for \textit{Orion-LD}), we required LLMs with rather large context windows. For our choice of models, we also wanted them to be of different authors but to have similar quality. Multiple benchmarks exist for comparing the performance of LLMs across different tasks, but we opted to look at the model pricing on OpenRouter as an indicator of quality. With these requirements in mind, we chose these three models: google/gemini-2.5-flash, openai/gpt-4.1-mini, and qwen/qwen3.6-flash, all costing some tens of cents per million input tokens, which was within our budget limitations.

Most of the repositories we have analyzed have documentation in the form of Markdown files. We concatenated these files to form a single big Markdown file for each repository. The exception was \textit{Snap4City}, for which the documentation is not in the repository itself but on its official website, where a comprehensive "Technical Overview" PDF document of 189 pages can be found. We chose that document to represent the documentation of the platform in our analysis. We converted this document to Markdown, using \href{https://github.com/datalab-to/marker}{Marker}. Since different LLMs can potentially parse PDFs in different ways, this ensures that all models are looking at the same input.

The prompt instructs the models to reply "Yes" only when the answer to a question is unambiguous. The models are also told to use the provided documentation to answer the questions, restricting their usage of external knowledge to the conceptual knowledge that is necessary to understand the questions and the documentation, without relying on factual knowledge they may possess about the analyzed platforms. Besides answering each question with Yes/No, the models are told to provide a short text explaining their decisions, along with the sections in the documentation that support them. This is an instance of Chain-of-Thought (CoT) reasoning, a prompt engineering technique that aims to improve the quality of the responses for complex tasks. In our context, it also allows us to more easily check the correctness of "Yes" answers, as a human can jump straight to the referenced sections and decide whether they contain valid evidence for a positive answer. Negative answers keep being more challenging to double-check, as the models cannot usually prove a negative by referencing a section.

\begin{table}
    \centering
    \small
    \caption{Docum. quality scores and inter-rater agreement}
    \label{tab:doc}
    
    \begin{tabular}{
        l
        S[table-format=2.2]
        S[table-format=2.0]
        S[table-format=1.3]
    }
        \toprule
        \textbf{Platform} &
        {\textbf{Mean}} &
        {\textbf{Majority Vote}} &
        {\textbf{Fleiss'$\kappa$}} \\
        \midrule
        Orion-LD     & 12.67 & 13 & 0.734 \\
        Snap4City    & 11.33 & 13 & 0.285 \\
        CIVITAS/CORE & 10.67 & 11 & 0.310 \\
        FogFlow      & 10.33 & 12 & 0.470 \\
        OpenRemote   &  8.67 &  7 & 0.036 \\
        UDP          &  8.67 &  9 & 0.332 \\
        \bottomrule
    \end{tabular}
\end{table}

For each platform, we determined the number of positive answers in two ways: the arithmetic mean and the majority vote (see Table~\ref{tab:doc}). We also analyzed the inter-rater agreement between the different LLMs at two levels: per platform and globally. We used Fleiss' $\kappa$~\cite{Fleiss1971} in both cases. The global Fleiss' $\kappa$ score is 0.368, which is considered a `fair' level of agreement~\cite{Landis1977}. The per-platform Fleiss' $\kappa$ can be seen as a measure of how unambiguous the documentation is. The majority vote yielded the same score for \textit{Orion-LD} and \textit{Snap4City}. However, the former has much higher inter-rater agreement, thus giving us more confidence that this platform's documentation is of a relatively high quality with respect to the questions we have devised.

\begin{table*}
    \centering
    \small
    \caption{Security indicator (OpenSSF Scorecard), complexity indicator (CC/LOC), number of commits, platform size (KLOC), team activity (commits per KLOC), reported bugs, mean time to repair issues labeled as bugs (days)}
    \label{tab:metrics}
    \begin{tabular}{
        l
        S[table-format=1.1]
        S[table-format=1.2]
        S[table-format=4.0]
        S[table-format=4.0]
        S[table-format=3.0]
        S[table-format=2.1]
        S[table-format=3.0]
        S[table-format=3.0]
        }
        \toprule
        
        \textbf{Platform} &
        \textbf{Scorecard} &
        \textbf{CC/LoC}    &
        \textbf{Stars} &
        \textbf{Commits} &
        \textbf{KLOC} &
        \textbf{Team Activity} &
        \textbf{Bugs} &
        \textbf{MTTR} \\
        
        \midrule
        
        CIVITAS/CORE & 2.0 & 0.01 & 23 & 1049 & 25 & 42.0 & {--}  & {--} \\
        FogFlow      & 2.9 & 0.09 & 130 & 48 & 147 & 0.3 & {--}  & {--} \\
        OpenRemote   & 7.2 & 0.11 & {1.8k} & 630 & 266 & 2.4 & 191 & 184\\
        Orion-LD     & 4.4 & 0.14 & 64 & 1170 & 289 & 4.0 & 15 & 129 \\
        Snap4City    & 0.5 & 0.11 & 28 & 81 & 982 & 0.1 & {--}  & {--} \\
        UDP          & 2.4 & 0.00 & 5 & 307 & 94 & 3.3 & {--}  & {--}  \\
        
        \bottomrule
    \end{tabular}
\end{table*}

\subsection{Code complexity}

We used \href{https://github.com/aveloxis/aveloxis}{Aveloxis} to collect repository-level metrics from the GitHub and GitLab repositories of the selected platforms. The data presented in Tables~\ref{tab:metrics} and~\ref{tab:community} were computed over the two years from June~2024 to June~2026.

Aveloxis integrates into its collection pipeline another tool, named \href{https://github.com/boyter/scc}{scc} (Sloc Cloc and Code) that analyzes source code and calculates per-file metrics such as the different types of lines (e.g., lines of code, comments, blank lines) and cyclomatic complexity (CC)~\cite{McCabe1976}. To create a per-repository metric, we normalized the complexity by dividing the sum of CC across files by the sum of lines of code across files, giving us the CC/LoC metric in Table~\ref{tab:metrics}. \textit{CIVITAS/CORE} and \textit{UDP} have near-zero CC/LoC due to their nature of being primarily integrations of different tools rather than typical software applications written in a programming language. The other platforms have rather similar values between themselves, with \textit{Orion-LD} having the highest CC/LoC.

\subsection{Maintainability}

The mean time to repair (MTTR) a bug, as shown in Table~\ref{tab:metrics}, is a commonly used maintainability metric. For each issue labeled as a bug that was active during the time period, we counted the days it took to close the issue. We only considered bugs that were closed within the time period; many such bug issues have not been closed yet or were closed after the considered time period (71 bugs for \textit{OpenRemote} and 37 for \textit{Orion-LD}). We could not analyze \textit{CIVITAS/CORE} and \textit{UDP} issues due to a lack of permission for the GitLab API. As for \textit{FogFlow}, \textit{OpenRemote}, and \textit{Snap4City}, no issues labeled as bugs were active during the time period. While this could be read as an indication that no bugs existed in these projects, another possible interpretation is that the community was not engaged enough to report bugs on GitHub.

\subsection{Security}

Security is a crucial requirement that an SCP must fulfill~\cite{Goumopoulos2024}. Aveloxis integrates into its collection pipeline the OpenSSF (Open Source Security Foundation) \href{https://scorecard.dev/}{Scorecard}, which automatically scores a project across several ``checks'', such as the number of unfixed vulnerabilities, code review practices, whether the project declares its license, among others. Each check is scored from 0 to 10. Some of these checks are clearly more important than others, and their importance can be context-dependent. Aggregated scores can thus be calculated based on different sets of weights. For our analysis, we have used the weighting system described in the Scorecard's repository. The security checks performed by Scorecard encompass several health characteristics: vulnerability presence, security practices, dependency management, and organizational diversity. Table~\ref{tab:metrics} shows the Scorecard aggregated scores for the platforms under evaluation.

\subsection{Community maturity}

\begin{table*}[t]
    \centering
    \small
    \caption{Community participation metrics. The qualitative assessment summarizes the observed maturity of the contributor community based on the breadth of participation, distribution of development effort, and review activity.}
    \label{tab:community}
    
    \begin{tabular}{
        l
        >{\centering\arraybackslash}p{0.9cm}
        S[table-format=2.0]
        S[table-format=2.0]
        S[table-format=2.0]
        >{\raggedright\arraybackslash}p{8cm}
    }
        \toprule
        
        \textbf{Platform} &
        \textbf{Core} &
        \textbf{Committers} &
        \textbf{Issue Only} &
        \textbf{Reviewers} &
        \textbf{Qualitative Assessment} \\
        
        \midrule
        
        OpenRemote &
        9 &
        27 &
        14 &
        18 &
        Large and well-balanced contributor community with strong peer-review participation. \\
        
        Orion-LD &
        3 &
        9 &
        30 &
        3 &
        Small development team but broad external engagement through issue reporting. \\
        
        CIVITAS/CORE &
        11 &
        31 &
        {--}  &
        {--}  &
        Large contributor base; issue-reporting and review data unavailable due to GitLab API limitations. \\
        
        UDP &
        6 &
        15 &
        {--}  &
        {--} &
        Medium-sized contributor community; issue-reporting and review data unavailable due to GitLab API limitations. \\
        
        FogFlow &
        3 &
        4 &
        0 &
        0 &
        Small and highly concentrated contributor community with no observable external participation. \\
        
        Snap4City &
        1 &
        1 &
        0 &
        0 &
        Single institutional account authors commits; the repository likely acts as a public mirror of the internal workspace. \\
        
        \bottomrule
    \end{tabular}
\end{table*}

To characterize the community participation on each platform, we identified four complementary contributor roles: the core team, the committers, issue-only contributors (users whose only contribution was reporting issues), and reviewers involved in pull-request discussions. The counts for these roles, shown in Table \ref{tab:community}, provide a snapshot of how participation is distributed across different roles within each repository. The size of the core team was estimated using a Pareto-based approach, defined as the smallest set of contributors responsible for 90\% of all commits. This method follows the approach used in previous studies on open-source project health and contributor concentration \cite{Valiev2018}. Reviewers include all contributors who approved, requested changes, or commented on pull requests submitted by others; self-reviews were excluded. For repositories hosted on GitLab, the available API permissions did not allow us to retrieve issue participation data. Consequently, \textit{CIVITAS/CORE} and \textit{UDP} do not have values for issue-only contributors or reviewers, and these missing values should not be interpreted as zero participation.

Table~\ref{tab:community} summarizes the observed participation patterns. Rather than aggregating the metrics into a single numerical score, we performed a qualitative assessment of community maturity, considering three dimensions: breadth of participation (number of contributors), distribution of development effort (core team relative to committers), and peer-review activity (number of reviewers). According to this assessment, \textit{OpenRemote} exhibits the most mature and balanced contributor community, combining a relatively large developer base with substantial review participation and external engagement. \textit{Orion-LD} shows a smaller development team but a comparatively broad community of issue reporters, indicating active user involvement. \textit{CIVITAS/CORE} and \textit{UDP} have sizable contributor communities, although the absence of issue and review data prevents a complete assessment. \textit{FogFlow} presents a small and concentrated contributor structure, while \textit{Snap4City}'s repository features a single institutional account as the committer, likely meaning that the repository is a public mirror of an internal collaborative workspace in which multiple members of DISIT Lab participate. Based on the available evidence, the platforms can therefore be ordered by decreasing community maturity as represented in Table \ref{tab:community}. This ranking should be interpreted as an assessment of the observed contributor community, not of the overall governance model or total user population of each platform.


\section{Discussion}
\label{sec:discussion}

\subsection{Consolidation}
    
\begin{table*}[h]
    \centering
    \small
    \caption{Ranking of the platforms along the five previously discussed health dimensions}
    \label{tab:summary} 
    \begin{tabular}{
        l
        c
        c
        c
        c
        c
        c
    }
        \toprule
        \textbf{Platform} &
        \textbf{Docum. Quality} &
        \textbf{Code Complexity} &
        \textbf{Maintainability} &
        \textbf{Security} &
        \textbf{Comm. Maturity} &
        \textbf{AVERAGE RANK} \\
        \midrule
        Orion-LD     & 6 & 1 & 6 & 5 & 6 & 4.8 \\
        CIVITAS/CORE & 4 & 5 & 3 & 2 & 5 & 3.8 \\
        FogFlow      & 3 & 4 & 3 & 4 & 2 & 3.2 \\
        OpenRemote   & 2 & 3 & 1 & 6 & 4 & 3.2 \\
        UDP          & 1 & 6 & 3 & 3 & 3 & 3.2 \\
        Snap4City    & 5 & 3 & 3 & 1 & 1 & 2.6 \\
        \bottomrule
    \end{tabular}
\end{table*}

\begin{figure}[t]
    \centering
    \includegraphics[width=\columnwidth]{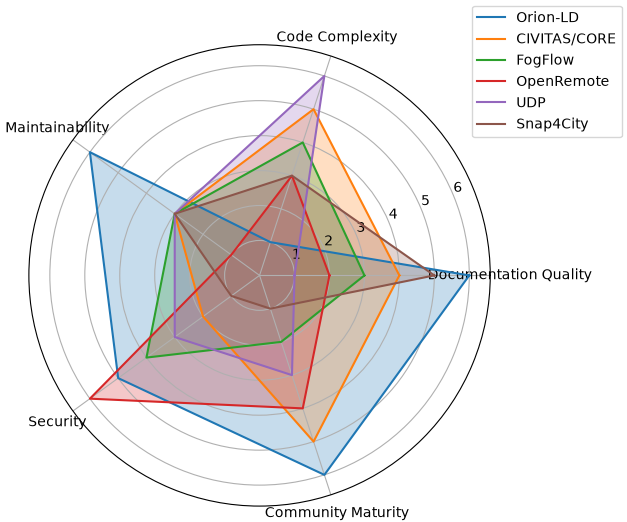}
    \caption{Platform ranks along the five health dimensions}
    \label{fig:kiviat}
\end{figure} 

Although our methodology evaluates a focused subset of health indicators based solely on public data, it highlights clear ecosystem trends. The consolidation of metrics across five of the evaluated health dimensions shows distinct profiles for each SCP in our analysis. Table~\ref{tab:summary} presents the comprehensive ranking, showing that \textit{Orion-LD} leads the evaluation with the highest average rank of 4.8. It achieved top-tier scores in documentation quality, maintainability, and community maturity. \textit{CIVITAS/CORE} follows as a strong alternative with a 3.8 average rank, driven by its performance in documentation and maintainability. A tied middle tier consists of \textit{FogFlow}, \textit{OpenRemote}, and \textit{UDP}, all securing an average rank of 3.2. While \textit{Snap4City} recorded the lowest overall average at 2.6, it is strong in documentation quality, and its low rank in community maturity is due to the decision of attributing all of the commits to an institutional account, which does not mean that the project has one developer only. Figure~\ref{fig:kiviat} visually reinforces these findings by mapping the platform ranks along the five dimensions in a radar chart.

\subsection{Threats to validity}

The automated analysis of documentation quality via LLM is not guaranteed to be as accurate as manual analysis by humans. To mitigate this issue, we instructed the LLMs to provide reasoning for their decisions. The responses are publicly available as \href{https://zenodo.org/records/22646494}{supplementary material}. Because our methodology involved examining only the text, the ``Graphical Support'' quality aspect of the documentation could not be evaluated. Furthermore, the answers to some of the quality evaluation questions may be found in the omitted figures (e.g., architectural diagrams), which could lead to false negatives.

\section{Conclusion and Future Work}
\label{sec:conclusion_future_work}

In this paper, we presented several health characteristics of various open-source SCPs. Our goal was not to recommend the adoption of a specific SCP, but rather to provide an overview of the health of these platforms. Besides the brief descriptions in Subsection~\ref{subsec:overview}, we did not analyze the features of the different platforms, which are critical when deciding which one to adopt.

We searched for source code repositories because we wanted to conduct deeper analyses of open-source platforms instead of relying on proxies for information, such as academic literature, as has been done in related work. For instance, we could install these platforms in a test environment to evaluate characteristics such as installability and configurability. These characteristics are important with respect to the motivation outlined in Section~\ref{sec:introduction}.

As mentioned in Section~\ref{sec:related_work}, one possible approach to deciding which characteristics to prioritize in a health analysis is devising user stories. Given the motivation outlined in Section~\ref{sec:introduction}, our user stories could take the form of \textit{``As a program manager in a municipality, I want to evaluate [health characteristic], so that the risk of OSS adoption is mitigated.''} Ideally, we would interview municipal staff to collect these user stories rather than imagining the needs and concerns of these stakeholders ourselves.


\bibliographystyle{IEEEtranDOI}

\bibliography{root} 
	
\end{document}